\documentclass{ceab}  

\usepackage{epsfig}    
\usepackage{graphicx}   

\usepackage{ceabbib}     

\usepackage[T1]{fontenc}

\begin{document}

\title{STUDY OF SUNSPOT GROUP MORPHOLOGICAL VARIATIONS LEADING TO FLARING EVENTS}

\author{Kors\'os, M. B., T. Baranyi and A. Ludm\'any
\vspace{2mm}\\
\it Heliophysical Observatory, Hungarian Academy of Sciences, \\
\it H-4010 Debrecen, P.O.Box 30. Hungary}

\maketitle

\begin{abstract}

It is widely assumed that the most probable sites of flare occurrences are the locations of high horizontal magnetic field gradients in the active regions. Instead of magnetograms the present work checks this assumption by using sunspot data, the targeted phenomenon is the pre-flare behaviour of the strong horizontal gradients of the magnetic field at the location of the flare. The empirical basis of the work is the SDD (SOHO/MDI-Debrecen sunspot Data) sunspot catalogue. Case studies of two active regions and five X-flares have been carried out to find possible candidates for pre-flare signatures. It has been found that the following properties of the temporal variations of horizontal magnetic field gradient are promising for flare forecast: the speed of its growth, its maximal value, its decrease after the maximum until the flare and the rate of its fluctuation.

\end{abstract}


\def\gore{Pre\,--\,flare variations}

\section{Introduction}

After a significant development in the flare forecast methods in the past decade the reliable prediction of the eruptive events in solar active regions remained an important and challenging task. The motivation is strong because of the high geoefficiency of flares and CMEs. Barnes and Leka (2008) as well as Bloomfield et al. (2012) published evaluation methods to assess their reliability. The empirical basis of the methodological development is the substantial progress in the solar observations of high temporal and spatial resolution providing an increasing amount of information about the active region magnetic fields. 

The released energy in flares is provided by the free magnetic energy, its source is the nonpotential component of the active region magnetic field (Priest and Forbes, 2002). The potential field does not contain electric currents, such magnetic configurations are not flare-productive. The energy release happens in the domains of large coronal currents where magnetic reconnections may take place. The large current layers cannot be observed directly, their detection would require the detailed knowledge of the magnetic field structure by extrapolating the data of surface magnetic fields to the corona, but this is an extremely difficult task. The recent attempts therefore focus on directly observable signatures of nonpotentiality to find diagnostically reliable pre-flare properties. 

The most promising directly observable feature is the inversion line separating the two areas of opposite polarities in the active regions. Schrijver (2007) showed on a large sample that the occurrences of intensive flares are connected to inversion lines of high magnetic gradient. Mason and Hoeksema (2010) defined a parameter named gradient-weighted inversion line length (GWILL) and found that it increased significantly prior to flares. Further proposed proxies of nonpotentiality: long inversion line with strong gradient and shear (Cui et al. 2007),  magnetic energy dissipation (Abramenko et al. 2009). The total magnetic flux (proposed by Song et al. 2009) and  free energy (Wang, 2007) are connected in a diagram called 'main sequence of explosive solar active regions' by Falconer et al. (2009). Further proposed pre-flare signatures are: magnitude of injected helicity in the active region (Labonte et al., 2007), fractality of the surface magnetic field  (Georgoulis, 2005; Criscuoli et al. 2009). 

All of the above mentioned methods are based on magnetograms. On the other hand, they mostly focus on characteristic  magnetic configurations and parameter values of pre-flare states, only a few works investigate dynamic phenomena prior to the flares as e.g. Liu et al. (2008). We want to study the series of events leading to flares at the location of the observed flares. For this purpose we use the data of sunspots which are discrete entities instead of the maps of continuous magnetic field distributions but they are locations of high flux densities.

Our present aim is to analyse the pre-flare values and the behaviour of the horizontal magnetic field gradient in the area of the flare in order to find indicative values of this behaviour during two or three days prior to the flare onset. The task is to carry out case studies about the precursors of some selected energetic flare events.

\section{Data and analysis}

The most suitable dataset for this investigation is the SDD (SOHO/MDI-Debrecen sunspot Data), the most detailed sunspot dataset for the years of MDI operations, 1996\,--\,2010. It contains the data of positions, areas and magnetic fields for all observable sunspots and sunspot groups on a 1.5 hourly basis (Gy\H ori et al. 2011). This detailedness allows to follow the evolution of the internal magnetic configuration of the active regions in high spatial and temporal resolution. 

Several investigations focus on the horizontal gradient of the line-of-sight (LOS) magnetic field to analyze its role in the flare onset, they are based on the magnetic field distributions registered in magnetograms. Our approach is based on sunspots so it should be checked as to how reliably they represent the high flux density component of the active region magnetic field. The SDD contains the mean field strengths of the sunspot umbrae and penumbrae. Figure 1 shows the dependence of the magnetic field strength on the area of the umbrae at different distances from the central meridian. This dependence means that the LOS magnetic field data are not suitable directly to follow the variations of the magnetic configuration during the passage of the active region through the solar disc, it should be replaced by a more reliable quantity.

\begin{figure}[h]
  \begin{center}
   \epsfig{file=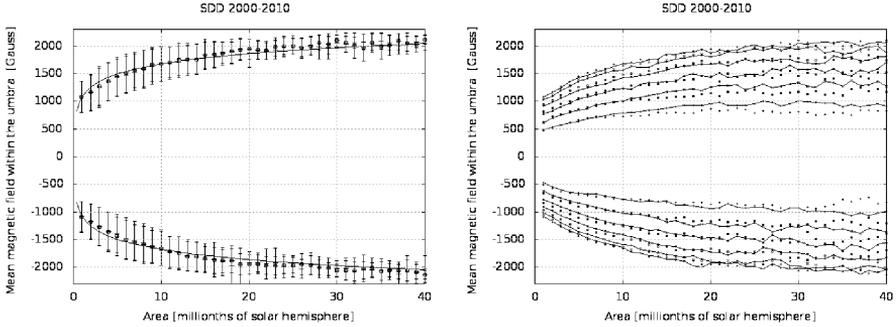,width=12cm}
  \end{center}
  
  \caption{Left panel: dependence of the magnetic field strength on the area of the umbrae at the center of the disc for both polarities. The eastern and western positions are distinguished but they are practically identical. The fitted curve is given in Equation (1). Right panel: the same dependences in six longitudinal bins from $\pm(0-10)^{\circ}$ (outer curves) to $\pm(50-60)^{\circ}$ (internal curves). The error bars are not indicated, their mean value is comparable to those plotted in the left panel.}
\end{figure} 

Figure 2 shows another statistics, the center-limb variation of observed sunspot size distributions. The sunspot umbrae of different sizes are also not equally observable on the solar disc. The SDD contains both the projected (observed) and the corrected areas of umbrae but this does not help to the smaller spots which become gradually unobservable toward the limb. However, the statistics of spots larger than the corrected area of 6-7 MSH (Millionths of Solar Hemisphere) is nearly independent on the distance of disc center. Thus it seems reasonable to represent the amount of flux by the umbral area which is more independent on the disc position. Of course, the umbrae of smallest size can only be considered within the $\pm30^{\circ}$ environment of the disc center.

\begin{figure}[h]
  \begin{center}
   \epsfig{file=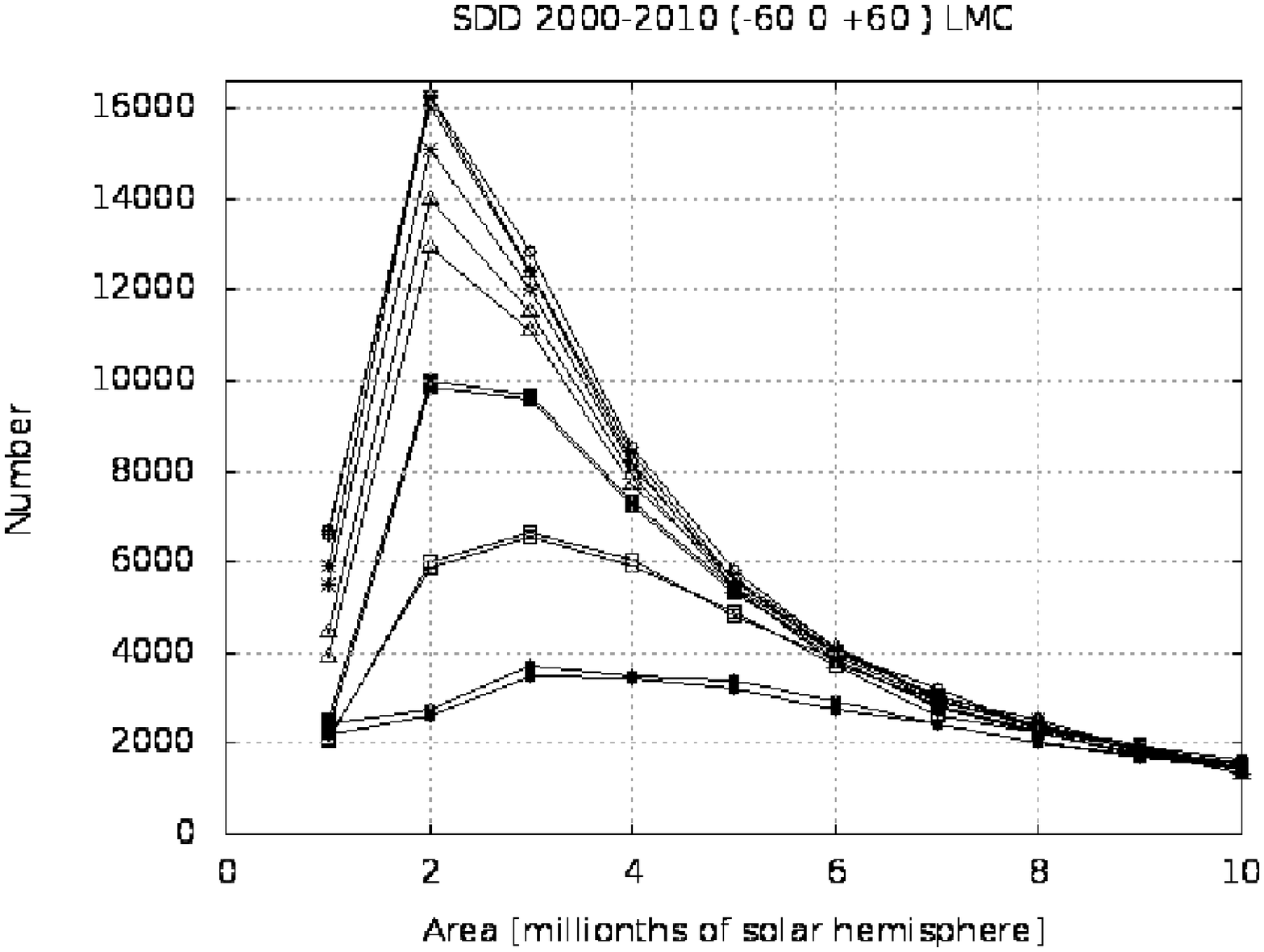,width=6cm}
  \end{center}
  \caption{Dependence of the numbers of umbrae on their sizes and positions. The symbols mean longitudinal bins measured from the central meridian in degrees: open circles: 0-10, stars: 10-20, triangles: 20-30, black squares: 30-40, open squares: 40-50, black dots: 50-60. The double curves refer to the eastern and western halves of the disc, they are practically identical.}
\end{figure} 

The targeted physical quantity is the horizontal gradient of the magnetic field between two areas of opposite magnetic polarities. The magnetic field will be represented by the total amount of magnetic flux enclosed within the umbra. The following logarithmic function has been fitted on the variation of the magnetic field measured in umbrae of different areas at the center of the solar disc (within $\pm10^{\circ}$ from the central meridian, see Figure 1):

\begin{equation}
      B \equiv f(A) = K_{1}*ln(A) + K_{2}
\,,
   \end{equation}

where $K_{1}$ = 264.5575 gauss and $K_{2}$ = 1067.0135 gauss. The same expression holds with negative signs for negative magnetic fields. This has been multiplied by the umbral area, this gives the amount of magnetic flux within the umbra. Thus the horizontal magnetic field gradient between two umbrae of opposite polarity is represented by the gradient of the magnetic flux:

\begin{equation}
      G_{M} = \left| \frac{\Delta(f(A)A)}{d} \right|
\,,
   \end{equation}

The measured values will be converted to SI units: Wb/m. We consider this quantity a possible proxy of nonpotentiality at the photospheric level. 

For case studies some energetic events have been selected to follow the development of the magnetic field gradient and to test the viability of the suggested approach. Figure 3 shows the NOAA 9393 active region on 29 March 2001, its white-ligth appearance, the magnetogram and (in the middle) the view of the sunspot group reconstructed from the SDD data. 


Three areas of the active region are separated by boxes in the middle panel of Figure 3. There were no significant eruptive events in boxes 1 and 3, but three flares occured in box 2, namely: X1.7 on 29 March at 09:57UT, X1.4 and X20 on 2 April at 10:04UT and 10:51UT respectively. Figure 4 shows the developments of the horizontal magnetic flux gradients in each box between those sunspot pairs of opposite polarity where this gradient was the highest.

It is conspicuous that the variation of magnetic flux gradient is the most significant in box 2 where the flares occured. The first flare was preceded by a steep rise of the gradient during about 1.5 days, then a slight decrease during another 1.5 days with a fluctuation of about $2.1\cdot10^{5}$ Wb/m, then happened the flare. The fluctuation is represented by the standard error of the estimate. Thereafter another, less steep and longer, rise can be observed and another, steeper and shorter, decrease with significant fluctuations and these variations lead to two X-flares. The variations detected in boxes 1 and 3 as two counterexamples are suitable to check the significance of this behaviour. It is remarkable that both areas are fairly quiet.

\begin{figure}[h]
  \begin{center}
  \epsfig{file=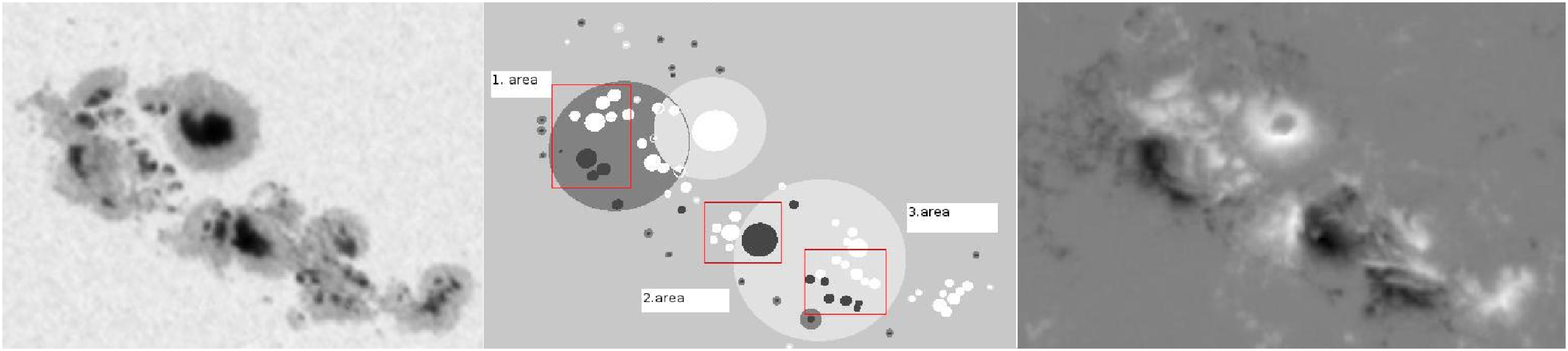,width=12cm}
   \epsfig{file=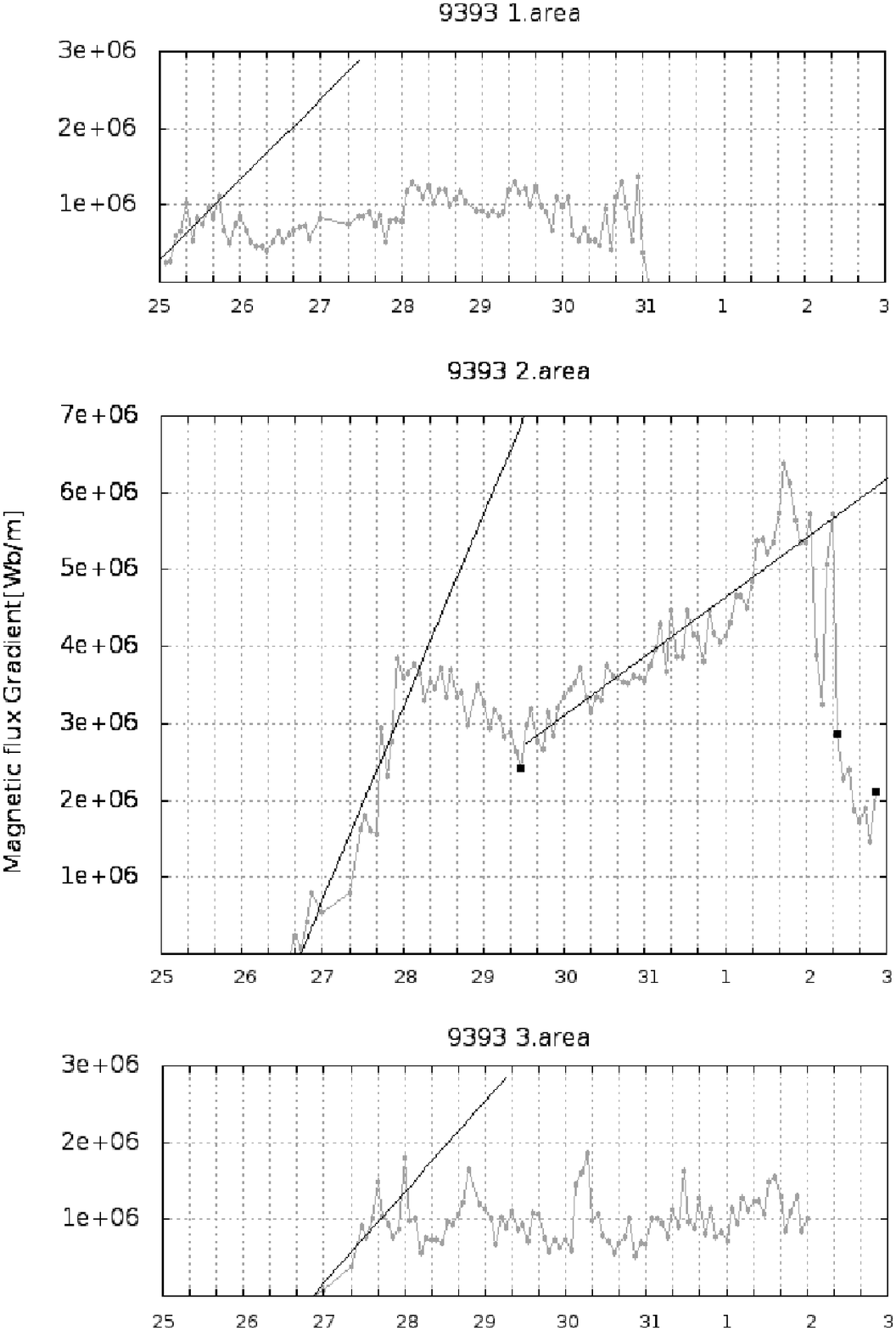,width=8cm}
   
  \end{center}
  \caption{Upper panel:White-light view of the NOAA 9393 active region on 29 March 2001 (left panel), its magnetogram (right panel) and the schematic view of the sunspot group reconstructed from the SDD (middle panel), the umbrae are represented by circles of corresponding area, the polarities are shown by colors.
  Lower panel:Variation of the horizontal magnetic flux gradient in the three boxes indicated in Figure 3 between 25 March and 3 April, 2001. Black squares show the times of flares.}
\end{figure}

In spite of the two similar variation patterns in box 2 one should be precautious to draw any conclusions about the pre-flare behaviour of the magnetic flux gradient but the diagrams are encouraging to scrutinize another case, the NOAA 10314 active region, see Figure 5. 
\begin{figure}[h]
  \begin{center}
   \epsfig{file=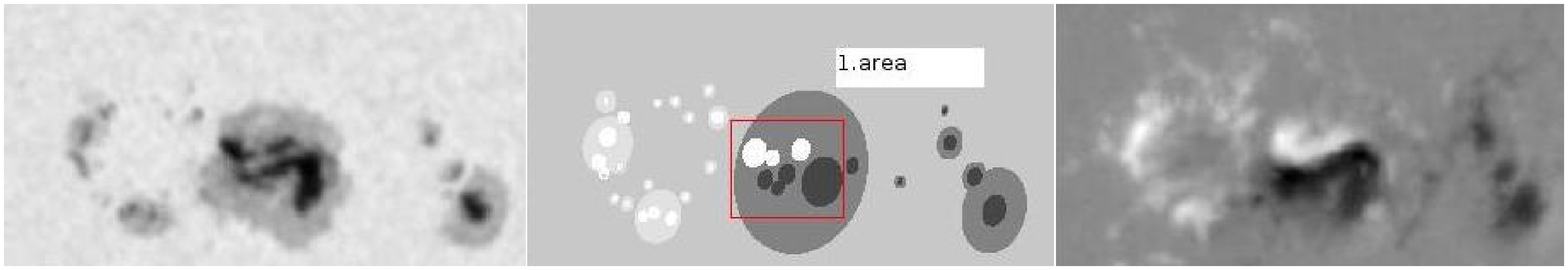,width=12cm}
   \epsfig{file=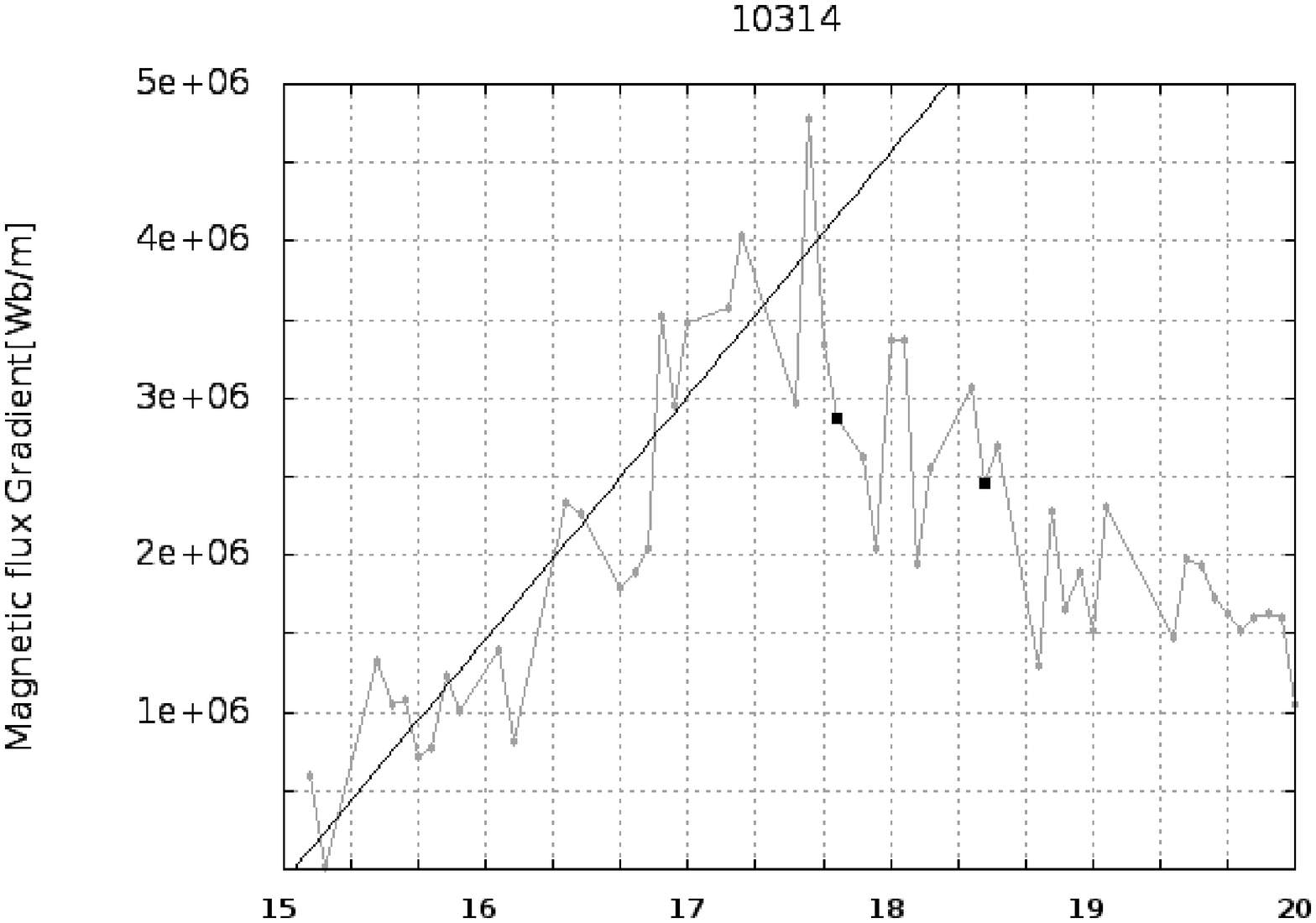,width=8cm} 
  \end{center}
  \caption{Upper panel, SOHO/MDI observations: white light image (left box), magnetogram (right box), and the view reconstructed from the SDD (middle box) of active region NOAA 10314 on 17 March 2003. Lower panel: time profile of the magnetic gradient within the "1.area" indicated in the middle box of upper panel.}
\end{figure} 

The upper panel of Figure 5 shows the different views of the active region on 17 March, 2003 like Figure 3, the lower panel shows the variation of the magnetic flux gradient and the onsets of  X1.5 flare on 17 March, at 18:50 UT and the X1.5 flare on 18 March at 11:51 UT. The common properties with the pre-flare variations in Figure 4 are: the steep rise of the gradient during more than 2 days, meanwhile a significant fluctuation of the gradient, and that the onsets of the flares are somewhat later than the maximum value after some decrease of the gradient. These properties can also be given quantitatively, the plots contain linear fits to the developing phases.

\begin{table}
\begin{tabular}{llcccc}
   \hline
date, time       & strength & growth                & max.               & fluct.               & time        \\
 &         &  rate                 & Wb/m               & Wb/m                 & after       \\  
           &         & Wb/m$\cdot$day        &                   &                      & max.        \\    
  \hline
2001.03.29 09:57 &   X1.7   & $1.6\cdot10^{5}$      & $3.8\cdot10^{6}$   & $2.1\cdot10^{5}$     & 35h 34m     \\
2001.04.02 10:04 &   X1.4   & $1.7\cdot10^{4}$      & $6.4\cdot10^{6}$   & $9.6\cdot10^{5}$     & 16h 28m     \\
2001.04.02 21:51 &   X20    & $1.7\cdot10^{4}$      & $6.4\cdot10^{6}$   & $2.9\cdot10^{5}$     & 29h 51m     \\
2003.03.17 18:50 &   X1.5   & $3.4\cdot10^{4}$      & $4.8\cdot10^{6}$   & $3.9\cdot10^{5}$     & 04h 27m     \\
2003.03.18 11:51 &   X1.5   & $3.4\cdot10^{4}$      & $4.8\cdot10^{6}$   & $6.3\cdot10^{5}$     & 18h 56m     \\
  \hline
\end{tabular}
\caption{Properties of the pre-flare behaviour of the horizontal magnetic field gradient at the location of the flares under study.}
\end{table}

\section{Discussions}

The published flare forecast methods use different properties of the active region magnetic fields to assess the probability of flare onset (Schrijver, 2009) and the validity of their predicting capabilities may also be different. For example, a gamma-configuration can always be expected to be flare-productive, this can be established even by studying a single observation. However, a more detailed tracking of the precursors is necessary if one has to enhance the information content of the forecast by including the expected time and strength of the flare. The present work is our first attempt to identify diagnostically promising features of the pre-flare dynamics. 

The proposed characteristics of the horizontal magnetic flux gradient are as follows: i) a steep rise during 1.5--2 days;  ii) maximum at about  $5\cdot10^{6}$ Wb/m;  iii) after the maximum a decrease for 0.5-1.5 days;  iv) during the decrease fluctuation of about $(5\pm2)\cdot10^{5}$ Wb/m.

The pre-flare quick growth of the flux gradient is not surprising, this is the build-up of the nonpotential component of the field. What is more intriguing, the decrease after the maximum until the onset of the flare. This may correspond to the theoretical result obtained by Kusano et al (2012) that the reconnection on a vertical current sheet is caused by the diverging flows that remove magnetic flux and plasma from the reconnection site. Yamada (1999) also mentions this process as a "pull" mode in laboratory experiments on reconnection. Its measured values are between $(3-6)\cdot10^{5}$ Wb/m  during the decrease prior to the X-flares whereas it is  $7.5\cdot10^{4}$ Wb/m in the quiet region No.1 of active region NOAA 9393 (Figure 4, top panel). If its role is really relevant, it may yield a clue for the triggering of the flares as a short timescale disturbance. It may be responsible for the quasi-periodic excitement of the current layer which finally enforces the reconnection event.

The presented signatures of pre-flare behaviour should be regarded as preliminary pieces of information for a more detailed, statistical analysis focusing on these specific characteristics. The aim is to establish a possible flare forecast method by close track of the horizontal magnetic field gradient variation.

\section*{Acknowledgements} 
The research leading to these results has received funding from the European Community's Seventh Framework Programme (FP7/2012-2015) under grant agreement No. 284461.


\section*{References}
\begin{itemize}
\small
\itemsep -2pt
\itemindent -20pt

\item[] Abramenko, V. I., Yurchyshyn, V. B., Wang, H., Spirock, T. J., Goode, P. R., 2003, {\it \apj}, 597, 1135-1144. 

\item[] Barnes, G., Leka, K. D., 2008, {\it \apj}, 688, L107. 

\item[] Bloomfield, D. S., Higgins, P. A., McAteer, R. T. J., Gallagher, P. T., 2012, {\it \apj}, 747, L41. 

\item[] Criscuoli, S., Romano, P., Giorgi, F., Zuccarello, F., 2009, {\it \aap},  506, 1429-1436.  

\item[] Cui, Y., Li, R., Wang, H., He, H., 2007, {\it \solphys}, 242, 1-8 

\item[] Falconer, D. A., Moore, R. L., Gary, G. A., Adams, M., 2009, {\it \apj}, 700, L166-L169 

\item[] Georgoulis, M. K., 2005, {\it \solphys}, 228, 5-27. 

\item[] Gy\H ori, L., Baranyi, T., \& Ludm\'any, A. 2011, {\it IAU Symp.} 273, 403. \\ see: http://fenyi.solarobs.unideb.hu/DPD/index.html

\item[] Kusano, K., Bamba, Y., Yamamoto, T. T., Iida, Y., Toriumi, S., Asai, A., 2012, {\it \apj}, 760, Issue 1, article id. 31, 9 pp. 

\item[] LaBonte, B. J., Georgoulis, M. K., Rust, D. M., 2007, {\it \apj}, 671, Issue 1, pp. 955-963. 

\item[] Liu, J., Zhang, Y., Zhang, H., 2008, {\it \solphys}, 248, 67-84. 

\item[] Mason, J. P., Hoeksema, J. T., 2010, {\it \apj}, 723, 634-640. 

\item[] Priest, E. R., Forbes, T. G., 2002, {\it AARv}, 10, 313-377. 

\item[] Schrijver, C., 2007, {\it \apj},  655, L117-L120. 

\item[] Schrijver, C. J., 2009, {\it AdSpR}, 43, 739-755. 

\item[] Song, H., Tan, C., Jing, J., Wang, H., Yurchyshyn, V., Abramenko, V., 2009, {\it \solphys}, 254, 101-125. 

\item[] Wang, H., 2007, {\it ASPC}, 369, 449. 

\item[] Yamada, M., 1999, {\it JGR}, 104, 14529-14542.

\end{itemize}

\end{document}